\documentclass[5p,twocolumn]{elsarticle}

\usepackage{graphicx}
\usepackage{subfigure}
\usepackage{longtable}
\usepackage{graphicx,wasysym}
\usepackage{longtable}
\usepackage{natbib}
\bibpunct{(}{)}{;}{a}{}{,}
\usepackage{amssymb}
\usepackage{bm}
\usepackage{enumitem}
\usepackage{epsfig}

\newcommand{\postscript}[2]{\setlength{\epsfxsize}{#2\hsize}
   \centerline{\epsfbox{#1}}}

\usepackage[usenames]{color}

\usepackage[figuresright]{rotating}
\usepackage{lscape}

\journal{JHEAP}

\begin{document}
\begin{frontmatter}

\title{Spatial curvature sensitivity to local $H_0$ from the Cepheid distance ladder}

\author{Ella
  Zuckerman$^a$ and Luis A. Anchordoqui$^{b,c,d}$}

\address[packer]{Packer Collegiate Institute, Brooklyn, NY 11201, USA
}
\address[lcuny]{Department of Physics and Astronomy,  Lehman College, City University of New York, NY 10468, USA}
\address[gccuny]{Department of Physics,
 Graduate Center, City University of New York,  NY 10016, USA}
\address[amnh]{Department of Astrophysics,
 American Museum of Natural History, NY 10024, USA}

\begin{abstract}
Over the last few years, low- and high-redshift observations set off a
tension in the measurement of the present-day expansion rate, $H_0$. Adding to the riddle, observational data from the Planck
mission point to a $3.4\sigma$ evidence for a closed
universe, further challenging the $\Lambda$CDM concordance model of
cosmology. Recently, a direct-observational test has been proposed to
discriminate effects of the spatial curvature in the cosmological
model. The test is based on the fundamental distance--flux--redshift
relation of the luminosity distance modulus, $\Delta \mu$. We reexamine
the outcomes of this test and show that achieving the required $\Delta
\mu$ sensitivity to discriminate among cosmological models is
materially far more challenging 
than previously thought. Armed with supernova
type Ia (SN Ia) data, calibrated using Cepheid measured distances, we
apply the test to archetypal spatially non-flat models that ameliorate
the $H_0$ tension and show that the $3\sigma$ contour of $\Delta \mu$
predicted by these models overlaps the
68\% CL SN Ia residuals with respect to $\Lambda$CDM. This implies
that the spatial curvature remains insensitive to local $H_0$ measurements from the Cepheid distance ladder.
\end{abstract}

\end{frontmatter}


\section{Introduction}

Almost a century after the expansion of the Universe was
established~\citep{Hubble:1929ig}, the Hubble constant,  which measures
its rate ($H_0~\equiv~100 \, h~{\rm km/s/Mpc}$), continues to encounter
challenging shortcomings. According to the latest observations, the
measured expansion rate~\citep{Riess:2020fzl,Anand:2021sum} is about 9\% faster
than predicted by observations of the cosmic microwave background
(CMB) on the
basis of the spatially flat $\Lambda$ cold dark matter (CDM) cosmological
model~\citep{Planck:2018vyg}. The statistical significance of this discrepancy is about $4.4\sigma$, which gives rise to the so-called $H_0$ tension~\citep{DiValentino:2020zio,Shah:2021onj}.

A plethora of models extending $\Lambda$CDM have been proposed to
ameliorate the $H_0$ tension; see e.g.~\citep{DiValentino:2021izs} for
a recent review. These models either reduce the size of the sound
horizon at recombination, modifying the expansion rate in the
early-universe, or else shift the matter--dark energy equality to
earlier times than it otherwise would in $\Lambda$CDM with new physics
in the post-recombination universe. Then, to keep the locations of the
peaks in the CMB angular power spectrum fixed, $H_0$ increases, 
diminishing the tension. A particular class of models of interest
herein are those in which the background geometry is not spatially
flat. These models are motivated by observational data from the Planck
mission~\citep{Planck:2018vyg}, which point to a $3.4\sigma$ evidence
for a closed
universe~\citep{DiValentino:2019qzk,Handley:2019tkm}. Recently, a
direct-observational test has been proposed to discriminate effects of
the spatial curvature in the cosmological
model~\citep{Shirokov:2020dwl}. The test pivots on variations of the
luminosity distance modulus, $\Delta \mu$. In this work we reexamine
the outcomes of this test and show that achieving the required $\Delta
\mu$ sensitivity to discriminate among cosmological models is more complex 
than previously thought. 

\section{Relative luminosity distance as a discriminator of space-curvature}

The classical distance-ladder approach to measure $H_0$ combines
Cepheid period-luminosity relations with absolute-distance
measurements to local anchors so as to calibrate distances to supernova
type Ia (SN Ia) host galaxies in the Hubble
flow. For each (homogeneous
and isotropic) cosmological
model, a key parameter of local $H_0$ measurements is the predicted distance modulus, which is given by
\begin{equation}
  \mu (z; \bm{\theta}) = 5 \log_{10} \left(\frac{D_L}{10~{\rm Mpc}}
  \right) + 25 \,,
\end{equation}
where $\bm{\theta}$ are the cosmological
parameters, 
\begin{equation}
  D_L (z) = \frac{c}{H_0} \frac{1+z}{\sqrt{|\Omega_k|}} \
   {\rm sinn} \left(\sqrt{|\Omega_k|} \int_0^z \frac{dz'}{\mathfrak{h}(z')} \right)\,,
 \end{equation}
is the luminosity distance, $\Omega_k$ is the present-day value of the
curvature density normalized to the critical density, $\mathfrak{h}(z) = H(z)/H_0$ is the normalized Hubble parameter, and  where ${\rm sinn} (x) = \sin (x), \ x, \ \sinh(x)$ for closed
($\Omega_k <0$), flat ($\Omega_k =0$), and open ($\Omega_k >0$)
universes~\citep{Dhawan:2020xmp}. For low-redshift probes, the contribution of the radiation density parameter $\Omega_r$ to $\mathfrak{h}(z)$ can be safely neglected, and thus
\begin{equation}
\hspace{-0.7cm} \mathfrak{h}(z) = \sqrt{\Omega_m (1+z)^3 + \Omega_{\rm DE} (1+z)^{3(1+w)}
        +\Omega_k (1 + z)^2},
\end{equation}
where $\Omega_m$ is the present day value of the nonrelativistic
matter density and $\Omega_{\rm DE}$ the corresponding dark energy
density parameter, with
$\Omega_{\rm DE} + \Omega_m + \Omega_k + \Omega_r = 1$. The 
scaling of $\Omega_{\rm DE}$ is characterized by the
``equation-of-state'' parameter $w \equiv p_{\rm DE} /\rho_{\rm DE}$,
the ratio of the spatially-homogeneous dark energy pressure to its
energy density.
 
In the spirit of~\citep{Shirokov:2020dwl}, we define the
relative luminosity distance, $D_{\rm rel}$,  as the ratio of the
luminosity distance in a given cosmological model to the luminosity
distance in the fixed spatially flat $\Lambda$CDM model,  
\begin{equation}
  D_{\rm rel} (z) = \frac{D_L(z)}{D_{L_{\Lambda{\rm CDM}}} (z)} = F(z;
    \Omega_k, \Omega_m,w) \,,
\end{equation}    
which is independent of $H_0$. To discriminate models featuring
$\Omega_k \neq 0$ from $\Omega_k =0$, we adopt the relative luminosity distance modulus, given by the difference between a given model of
luminosity distance modulus and the spatially flat $\Lambda$CDM
luminosity distance modulus,
\begin{equation}
  \Delta \mu (z; \Omega_k,\Omega_m,w) = 5 \log_{\rm 10} \left[ F(z;
  \Omega_k,\Omega_m,w) \right] \,.
\end{equation}

We specify the range of cosmological parameters to scan over using 
results of Markov chain Monte
Carlo analyses, which confront the growth of perturbations and of CMB fluctuations with experimental
data. These analyses, which have been carried out elsewhere~\citep{DiValentino:2020hov,Anchordoqui:2021gji}, use the publicly available Boltzmann solver CAMB~\citep{Lewis:1999bs} in
combination with the sampler CosmoMC~\citep{Lewis:2002ah,Lewis:2013hha}, and the following data sets: {\it (i)}~the CMB temperature and
  polarization angular power spectra {\it plikTTTEEE+lowl+lowE} from
  the Planck 2018 legacy
  release~\citep{Planck:2018vyg,Planck:2019nip}; {\it (ii)}~measurements of  baryon acoustic oscillations (BAO) from different galaxy surveys
  (6dFGS~\citep{Beutler:2011hx}, SDSS-MGS~\citep{Ross:2014qpa}, and BOSS
  DR12~\citep{BOSS:2016wmc}); {\it (iii)}~the 1048 SN type Ia
  data points of the Pantheon sample distributed in the redshift 
interval $0.01 < z <2.3$~\citep{Scolnic:2017caz}; {\it (iv)}~a gaussian
prior on the Hubble constant in agreement with  measurements
obtained by the SH0ES Collaboration~\citep{Riess:2019cxk,Riess:2020fzl}.

\begin{figure}[b]
    \postscript{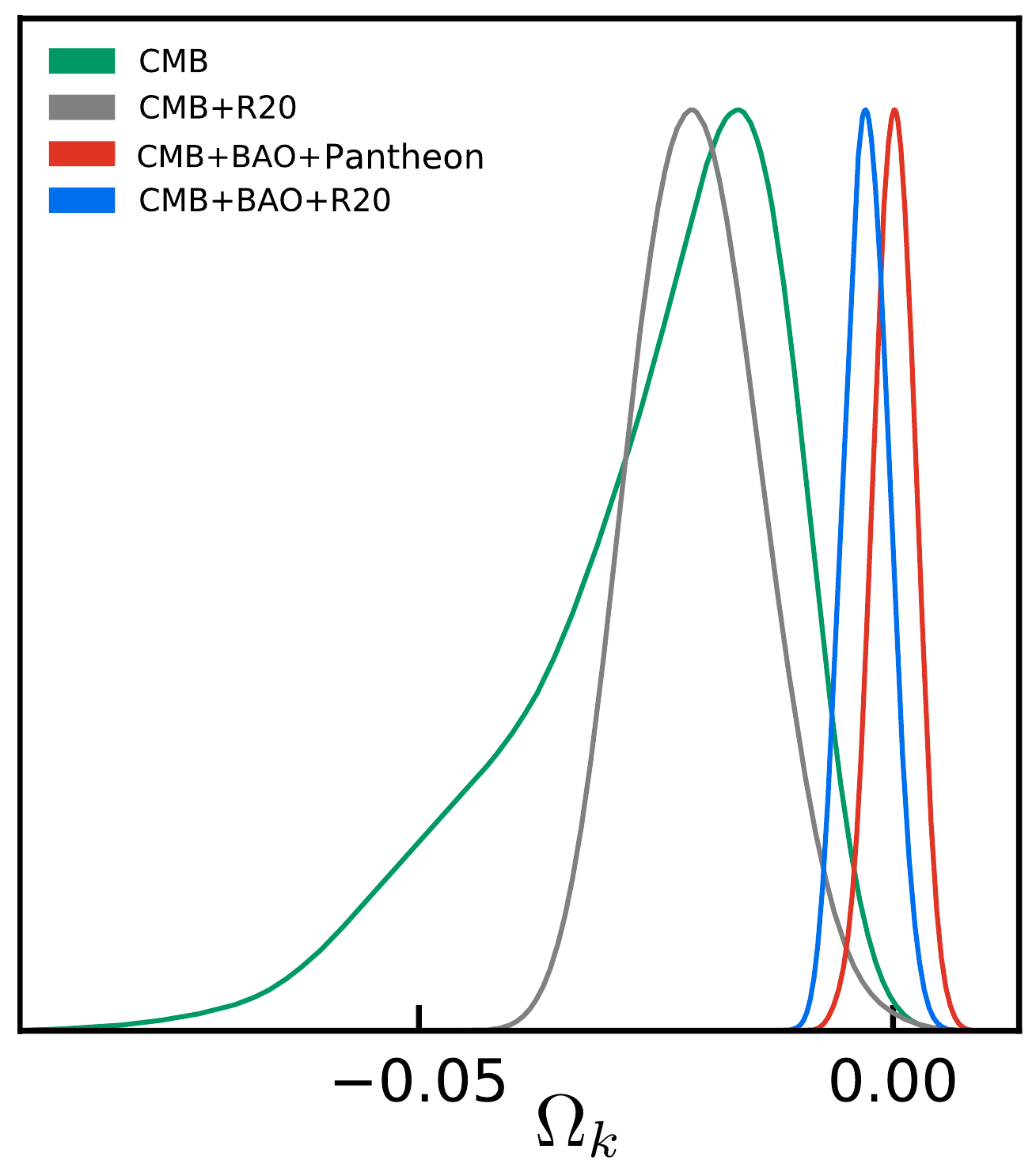}{0.85} 
    \caption{$\Omega_k$  one-dimensional
posterior distribution of ${\cal P}_1$; for details see~\citep{Anchordoqui:2021gji}.  \label{fig:1}}
\end{figure}

\begin{figure*}[tpb]
\begin{minipage}[t]{0.49\textwidth}
  \postscript{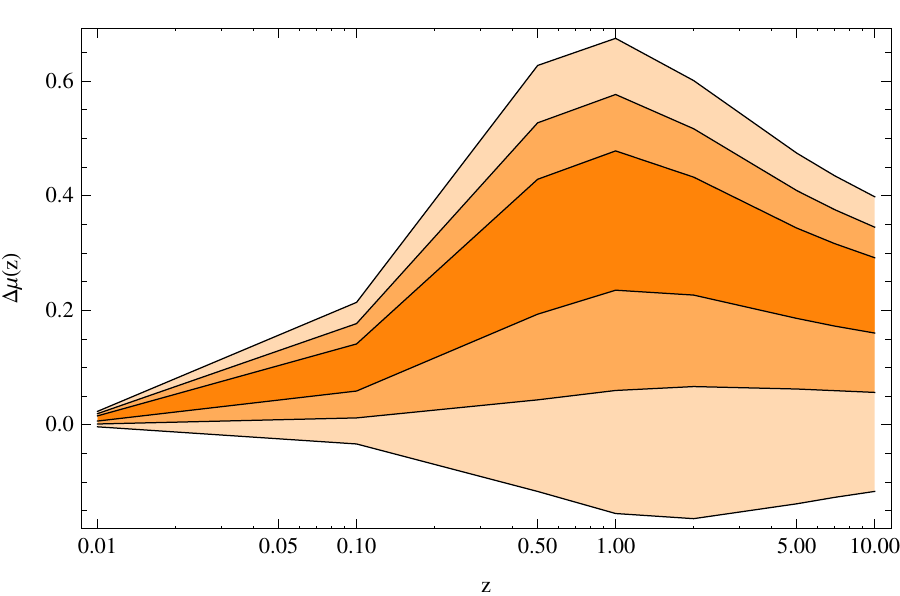}{0.95} 
\end{minipage}
\hfill
\begin{minipage}[t]{0.49\textwidth}
  \postscript{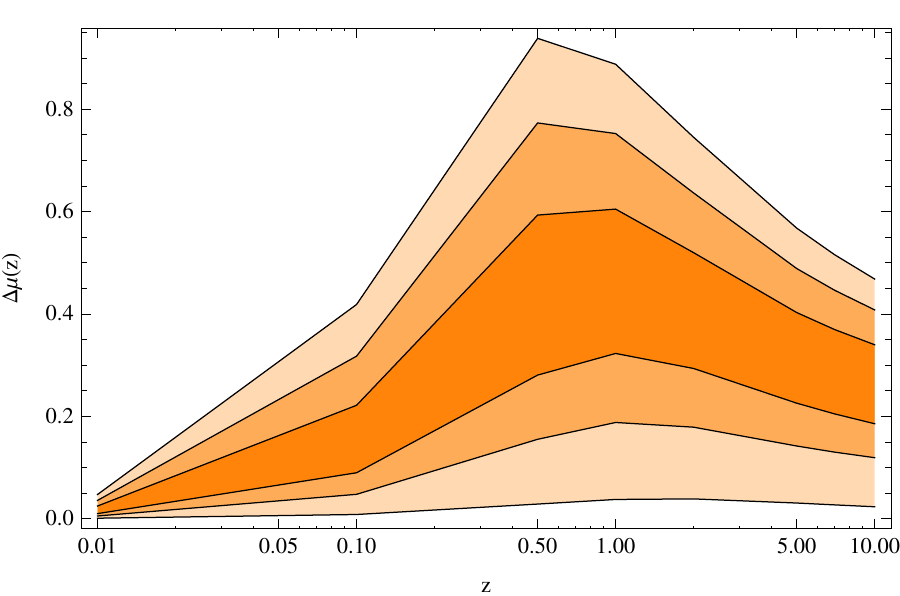}{0.95}
  \end{minipage}
\caption{Relative distance modulus $\Delta \mu (z)$ predicted by ${\cal P}_1$ (left) and ${\cal P}_2$ (right), for variations of $\Omega_k$, $w$, and $\Omega_m$
  within $1\sigma$, $2\sigma$, and $3\sigma.$  \label{fig:2}}
 \end{figure*}

We consider two representative cosmological models in which particular combinations of data
samples provide evidence for $\Omega_k \neq 0$, with a statistical
significance larger than $3\sigma$ and relax the $H_0$ tension:
\begin{itemize}[noitemsep,topsep=0pt]
\item A $\Lambda$CDM extension with three extra free parameters, which
  are $\Omega_k \neq 0$, $w \neq -1$,  and $N_{\rm eff}$; the latter characterizes the number of
  ``equivalent'' light neutrino species prior to
  recombination~\citep{Anchordoqui:2021gji}. In the $\Lambda$CDM model
  $N_{\rm eff} = 3.046$ for three families of massless (Standard
  Model) neutrinos~\citep{Mangano:2005cc}. The 9-parameter space is given by
\begin{eqnarray}
\hspace{-1.0cm} {\cal P}_1 & \equiv & \!\!\!\! \bigg\{\Omega_b h^2, \Omega_{\rm
  CDM} h^2, 100 \, \theta_{\rm MC}, \tau, n_s, \ln[10^{10} A_s],
                      \nonumber \\
  & & w,
\Omega_k, N_{\rm eff} \bigg\} \, ,
\end{eqnarray}
where  $\Omega_bh^2$ is the density
of baryons, $\Omega_{\rm CDM} h^2$ is the density of CDM, $\theta_{\rm
  MC}$ is the ratio of sound horizon to the angular diameter distance, $\tau$ denotes the reionization optical depth, $n_s$ is the scalar spectral index, and $A_s$ is the amplitude of the primordial scalar power spectrum.
\item A $\Lambda$CDM extension with four extra free parameters, which are $\Omega_k
\neq 0$, $w \neq -1$,
the sum of neutrino masses $\sum_i m_{\nu_i}$, and the running of the
spectral index of inflationary perturbations $\alpha_s$~\citep{DiValentino:2020hov}. In the
$\Lambda$CDM model a minimal $\sum_i m_{\nu_i} = 0.06~{\rm eV}$ is
assumed~\citep{Planck:2018vyg}. A $3\sigma$ indication for a negative running of $\alpha_s$
has been observed in the combined analysis of Planck, BAO, and
Lyman-$\alpha$ forest data~\citep{Palanque-Delabrouille:2019iyz},
while a positive value at more than $2.4\sigma$ has been measured by
the ACT Collaboration~\citep{ACT:2020gnv}; see also~\citep{Forconi:2021que}. The
10-parameter space is given by
\begin{eqnarray}
\hspace{-1cm} {\cal P}_2 & \equiv & \!\!\!\! \bigg\{\Omega_b h^2, \Omega_{\rm
  CDM} h^2, 100 \, \theta_{\rm MC}, \tau, n_s, \ln[10^{10} A_s],
                      \nonumber \\
  & &~~w,
\Omega_k, \alpha_s, \sum_i m_{\nu_i} \bigg\}\, .
\end{eqnarray}
\end{itemize}

To discriminate the models from $\Lambda$CDM, we require the predicted 
free parameters in the likelihood fit to be $3\sigma$ away from the
$\Lambda$CDM result for any particular combination of the data
samples yielding a solution of the $H_0$ tension. In Fig.~\ref{fig:1} we show the $\Omega_k$ one-dimensional
posterior distributions of ${\cal P}_1$. A larger than $3\sigma$ evidence of $\Omega_k \neq 0$ is only achieved for the combination of
CMB data with the SH0ES gaussian
prior on the Hubble constant (dubbed R20). The best-fit values of relevant cosmological
parameters to our analysis are: $\Omega_k =
-0.0207^{+0.0065}_{-0.0075}$, $w= -1.90^{+0.41}_{-0.25} $, and
$\Omega_m = 0.264^{+0.010}_{-0.012}$~\citep{Anchordoqui:2021gji}. For
${\cal P}_2$, both CMB data alone and the combination of CMB data with the SH0ES
prior give a larger than 
$3\sigma$ effect. However, when analyzing the CMB  data alone, the best-fit
gives a relative small evidence for the curvature density, $\Omega_k = -
0.074^{+0.058}_{-0.025}$, which exacerbates the Hubble tension: $H_0 =
 53^{+6}_{-16}~{\rm
   km/s/Mpc}$~\citep{DiValentino:2020hov}.\footnote{A point worth
   noting at this juncture is 
   that assuming a flat $\Lambda$CDM model the best-fit to extract cosmological parameters by the Planck Collaboration leads to
$H_0 = 67.27 \pm 0.60~{\rm km/s/Mpc}$ at 68\% CL~\citep{Planck:2018vyg}, whereas the SH0ES
Collaboration finds a larger value $H_0 = 73.2\pm  1.3~{\rm
  km/s/Mpc}$~\citep{Riess:2020fzl}.}  Therefore, we only consider for
our investigation the results of the
likelihood analysis to CMB data with the SH0ES prior. The best-fit
values of the 
relevant parameters are as follows: $\Omega_k =
-0.0192^{+0.0036}_{-0.0099}$, $w = -2.11^{+0.35}_{-0.77}$, $\Omega_m = 0.264^{+0.010}_{-0.013}$~\citep{DiValentino:2020hov}. 

\begin{figure}
    \postscript{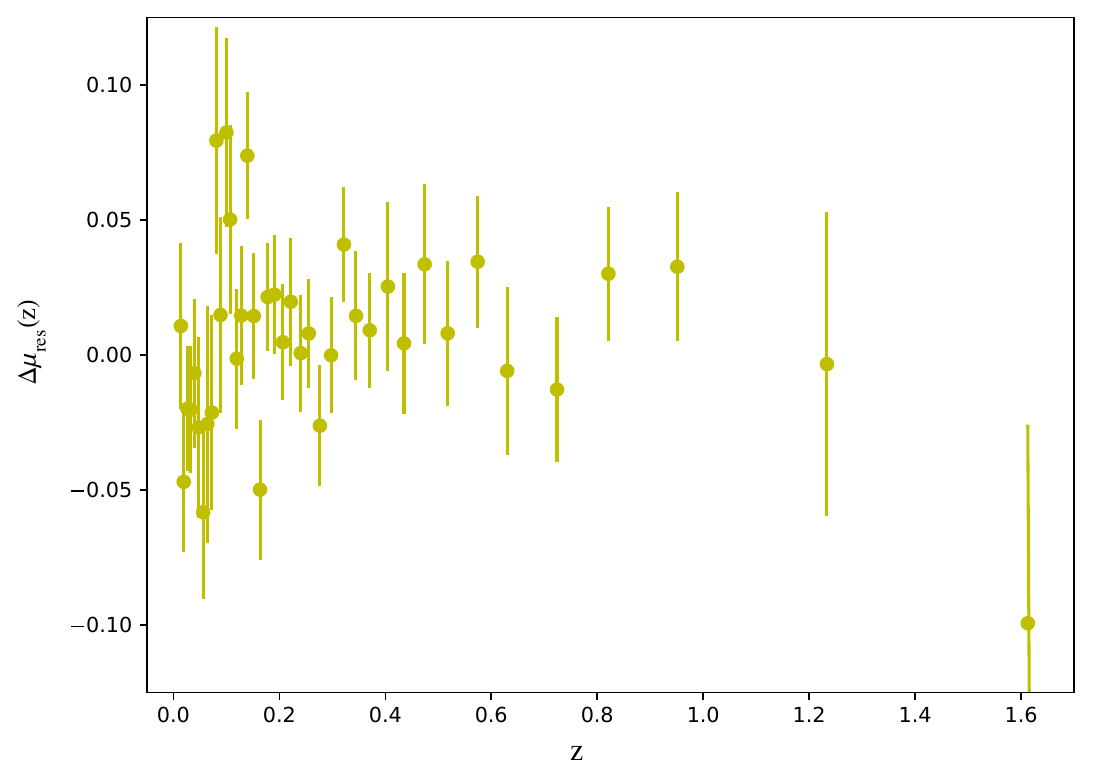}{0.95} 
\caption{Residuals for the SN Ia data plotted relative to the best fit
  $\Lambda$CDM model. Adapted from~\citep{Dhawan:2020xmp}.  \label{fig:3}}
\end{figure}

In Fig.~\ref{fig:2} we show the predicted relative distance modulus by
${\cal P}_1$ and ${\cal P}_2$ when varying the relevant parameters
between $1\sigma$, $2\sigma$, and $3\sigma$. We can see that for ${\cal P}_1$, the
predicted lower boundary of the $3\sigma$ region (corresponding to
$\Omega_m = 0.294$) is consistent with the $\Lambda$CDM prediction, whereas for ${\cal P}_2$, gives $\Delta
\mu \lesssim 10^{-2}$. Altogether, this challenges the feasibility of using  the
relative luminosity distance modulus to probe 
the spatial curvature. 

For ${\cal P}_2$, the predicted $\Delta \mu$ lower boundary of the $3\sigma$ region has a maximum
around  $z \sim 1$, which is currently probed by the Pantheon data 
sample. As an illustration, in Fig.~\ref{fig:3} we show the residuals for the SN Ia data plotted relative to the best fit
  $\Lambda$CDM model~\citep{Dhawan:2020xmp}. Now, a direct comparison of
Figs.~\ref{fig:2} and \ref{fig:3} demonstrates that the 68\% CL residuals of SN Ia
data relative to the best fit $\Lambda$CDM overlap
the predicted $\Delta \mu$ contours obtained from the $3\sigma$ variation of relevant model
parameters. This implies that for the models analyzed herein, the spatial curvature remains
insensitive to local $H_0$ measurements from the Cepheid distance ladder.

\section{Conclusions}

We have reexamined the idea of using the relative luminosity distance
modulus $\Delta \mu$ as a discriminator of
space-curvature~\citep{Shirokov:2020dwl}. We have shown that achieving the required $\Delta
\mu$ sensitivity to discriminate among cosmological models will be
vastly more complicated than once thought. An improvement in the
sensitivity to $\Delta \mu$ of future probes  must be accompanied by a reduction of the systematic uncertainties driving the determination of cosmological parameters.

\section*{Acknowledgements}

We thank Eleonora Di Valentino for valuable discussions. LAA is supported by the U.S. National Science Foundation Grant
PHY-2112527. 

\section*{Data availability}

We have used data from standard cosmological probes which are freely available.

\end{document}